\documentclass[12pt,preprint]{aastex}

\def\hhhp{H$_{3}^{+}$}
\def\hh{H$_2$}
\def\hhp{H$_{2}^{+}$}

\slugcomment{to be published in ApJ vol. 644, June 20, 2006}

\shortauthors{Geballe et al.}

\shorttitle{\hhhp\ and CO in IRAS 08572+3915 NW}

\begin{document} 

\title{The Interstellar Medium of IRAS
08572+3915 NW: \hhhp\ and Warm High Velocity CO\altaffilmark{1,2}}

\author{T. R. Geballe\altaffilmark{3}, M. Goto
\altaffilmark{4}, T. Usuda \altaffilmark{5} T. Oka\altaffilmark{6},
B. J. McCall\altaffilmark{7}}

\altaffiltext{1}{Based on data obtained at the United Kingdom Infrared
Telescope, which is operated by the Joint Astronomy Center on behalf of the
U. K. Particle Physics and Astronomy Research Council}

\altaffiltext{2}{Based on data collected at the Subaru
Telescope, which is operated by the National Astronomical Observatory of
Japan}

\altaffiltext{3}{Gemini Observatory, 670 N. A'ohoku Place, Hilo, HI
96720; tgeballe@gemini.edu}

\altaffiltext{4}{Max Planck Institute for Astronomy, Koenigstuhl 17,
D-69117 Heidelberg, Germany}

\altaffiltext{5}{Subaru Telescope, Astronomical National Observatory of
Japan, 650 N. A'ohoku Place, Hilo, HI 96720}

\altaffiltext{6}{Department of Astronomy and Astrophysics, Department of
Chemistry, 5735 S. Ellis Avenue; and Enrico Fermi Institute, University of
Chicago, Chicago, IL 60637}

\altaffiltext{7}{Department of Chemistry and Department of Astronomy,
University of Illinois at Urbana-Champaign, Urbana IL 61801}

\begin{abstract}

We confirm the first detection of the molecular ion \hhhp\ in an
extragalactic object, the highly obscured ultraluminous galaxy
IRAS~08572+3915~NW. We also have detected absorption lines of the
fundamental band of CO in this galaxy. The CO absorption consists of a
cold component close to the systemic velocity and warm, highly blueshifted
and redshifted components. The warm blueshifted component is remarkably
strong and broad and extends at least to -350 km~s$^{-1}$.  Some analogies
can be drawn between the \hhhp\ and cold CO in IRAS~08572+3915~NW and the
same species seen toward the Galactic center. The profiles of the warm CO
components are not those expected from a dusty torus of the type thought
to obscure active galactic nuclei. They are probably formed close to the
dust continuum surface near the buried and active nucleus and are
probably associated with an unusual and energetic event there.

\end{abstract} 
 

\section{Introduction} 

Ultraluminous infrared galaxies (ULIRGs) have been extensively studied at
all wavelengths from the radio to X-ray in order to probe the nature of
the energy source - starburst and/or active galactic nucleus (AGN).
Numerous investigators have concluded that ULIRGs are mainly found in
interacting systems \citep[e.g.,][and references therein]{san96, vei02}.  
However, in the many cases where the nuclear regions are heavily obscured
it is still unclear whether their huge luminosities are the result of
greatly enhanced nuclear star formation, or greatly enhanced accretion by
a central black hole, or both. Infrared spectroscopy potentially has an
important role to play in this investigation, as it can probe regions
close to the nucleus at high angular resolution and detect diagnostics
unique to each phenomenon, as well as discern the nature of the
interstellar medium of the ULIRG more distant from the nucleus. Recent
observations using this technique include the measurements of broad line
regions and coronal lines, which are signposts of AGNs, in several ULIRGs
\citep{vei99, mur01}, the detection of interstellar aromatic hydrocarbon
features and hydrogen recombination lines indicative of massive star
formation \citep{ima00, soi02, dan05} in a partially intersecting set of
ULIRGs, and most recently and perhaps most remarkably, the discovery of CO
in dense and warm gas close to the central luminosity source of the
z=0.327 ULIRG, IRAS F00183-7111 \citep{spo04}.

\hhhp, the highly reactive molecular ion upon which interstellar gas phase
chemistry is based \citep{her73, wat73}, is a potentially useful infrared
spectroscopic tool for studying the interstellar gas in distant galaxies.
It is important observationally for understanding both dark and diffuse
interstellar clouds \citep{geb00}. In both types of clouds it is produced
following cosmic ray ionization of \hh\ to \hhp, which quickly reacts with
\hh\ to form \hhhp.  The steady state abundance of \hhhp\ is very low,
because it is destroyed readily in dark clouds by reactions with neutral
molecules (principally CO) and atoms (mainly O) and even more readily in
diffuse clouds by dissociative recombination with electrons, which, due to
the photoionization of carbon, are much more abundant than in dark clouds.
The absorption strengths of \hhhp\ lines provide basic information on the
cloud dimensions and environment, in addition to temperature.  The
observed column density of \hhhp\ can directly yield the product of the
distance through the cloud and the cosmic ray ionization rate in the
cloud. This is because unlike most other molecules, the number density of
\hhhp\ in a cloud is a constant that depends only on the ionization rate
of H$_{2}$ and whether the cloud is dark or diffuse. This unusual property
of \hhhp\ comes about because its creation rate per unit volume is
proportional to the first power of the cloud density, rather than to the
square.

The detection of strong \hhhp\ absorption toward the center of the Galaxy
\citep{geb99} suggests that it also should be possible to detect \hhhp\ in
the interstellar medium of suitable external galaxies - those with
sufficiently bright and compact sources of infrared continuum radiation
and large column densities of interstellar molecular gas along their lines
of sight. One of the most promising galaxies for an \hhhp\ search is the
ULIRG, IRAS~08572+3915.  Images at several wavelengths of this interacting
pair of galaxies can be found in \citet{eva02}.  The 3.4~$\mu$m
interstellar hydrocarbon absorption observed by \citet{wri96} and
\citet{ima00} (also see Fig~1, this paper) in the northwest component of
this merging galaxy pair is deeper than that observed toward the Galactic
center. In the Galaxy this feature signifies extinction by dust in diffuse
clouds. The 10~$\mu$m silicate absorption toward IRAS~08572+3915~NW
\citep{dud97,spo06a} also is among the strongest detected. In addition
IRAS~08572+3915~NW possesses intense CO line emission at millimeter
wavelengths \citep{san89, eva02}. Thus extensive columns of diffuse and
dense gas exist in front of the nuclear infrared source of
IRAS~08572+3915~NW, which may contain a buried AGN \citep{dud97, ima00},
although this interpretation has been contested \citep{arr00}.

Here we report confirmation of the detection of \hhhp\ in IRAS~08572+3915~NW
reported by \citet{geb01}.  To gain additional information about the
interstellar medium of this galaxy we also obtained a spectrum of it in the
region of the fundamental band of carbon monoxide. This spectrum reveals both
a cold interstellar component, possibly associated with some of the gas
containing the detected \hhhp, and an unusual warm and dense component
superficially similar to that observed toward IRAS~F00183-7111 by
\citet{spo04}. Here, however, the CO band is resolved into individual lines
and the lines themselves are velocity-resolved, revealing remarkably broad
and high speed components.

A data log of the observations reported below is provided in Table~1. In
all cases data reduction was straightforward, involving removal of bad
frames, extraction of spectra from coadded 2D spectral images, wavelength
calibration using arc lamps and telluric absorption lines, despiking,
division by a telluric standard star observed at close to the same airmass
and reduced in a similar fashion, and flux calibration using the observed
or predicted flux density of the standard.

\section{Observations and Results} 

\subsection{\hhhp}

An initial search for \hhhp\ in IRAS~08572+3915~NW was made in 1998
December at the United Kingdom Infrared Telescope (UKIRT) with the use of
the facility spectrograph CGS4 \citep{mou90}, employed at a resolving
power of 1500 and covering 3.4-4.0~$\mu$m. No \hhhp\ lines were detected.
This spectrum is shown in Fig.~1, because it is a better quality spectrum
of the redshifted 3.4~$\mu$m hydrocarbon absorption feature in
IRAS~08572+3915~NW than those published to date \citep{ima00,mas04}. The
spectrum also illustrates the difficulty of detecting weak lines from the
ground at medium spectral resolution in the thermal infrared region.

In December 2000 a 3.82-3.98~$\mu$m spectrum of IRAS~08572+3915~NW was
obtained at UKIRT, with CGS4 configured to give a higher resolving power
of 6000. The spectrum covered the wavelength range of three lines of
\hhhp\ from the lowest lying ortho and para levels. These are the
R(1,1)$^{u}$ - R(1,0) doublet at 3.66808~$\mu$m and 3.66852~$\mu$m
(separated by 36 km~s$^{-1}$), whose mean wavelength is redshifted to
3.8822~$\mu$m and the R(1,1)$^{l}$ singlet at 3.71548~$\mu$m redshifted to
3.9321~$\mu$m. We use z=0.0583, which was determined from the CO pure
rotational 1-0 line \citep{eva02} and is probably accurate to better than
0.0001 (30~km~s$^{-1}$), as the nominal systemic redshift and express all
wavelengths {\it in vacuo}.  The relevant portion of the spectrum is shown
at the top of Fig.~2. Statistically significant absorption features were
observed at the expected wavelengths of the lines, indicating that the
molecular ion had been detected \citep{geb01}. The \hhhp\ search was
repeated in 2002 and 2004 at the Subaru Telescope using its Infrared
Camera and Spectrograph (IRCS) \citep{tok98, kob00} at resolving powers of
10,000 and 5,000.  These data, each summed in 0.0006~$\mu$m
($\sim$50~km~s$^{-1}$) wide bins to match the UKIRT point spacing, also
are shown in Fig.~2 and, except for the singlet in the 2004 spectrum, also
contain modest signal-to-noise-ratio detections of these lines. In the
2004 data the signal near the wavelength of the singlet is depressed as
expected, but it is also depressed at adjacent wavelengths.

The transmission spectrum of the earth's atmosphere in the
3.86-3.95~$\mu$m interval contains about fifty narrow and roughly evenly
spaced absorption lines from high altitude N$_2$O. In unratioed spectra
with the above binning the strongest of these are 20\% deep. Any effects
due to non-cancellation of these lines would result in systematic features
of comparable strengths at many wavelengths in the ratioed spectra. No
evidence for any such features is present in any of the individual
spectra. The broad and weak emission bumps centered near 3.908~$\mu$m in
the spectra are due to absorption by H~I 15-6 in the spectra of the A
dwarf telluric standards.

The mean of the three spectra is shown near the bottom of Fig.~2. In it
the detection of the doublet is convincing,whereas that of the singlet is
marginal (at about 3$\sigma$). Relative to the systemic velocity, the
centroids of the doublet and the singlet correspond to LSR velocities of
-50~$\pm$30~km~s$^{-1}$ and +10~$\pm$~50~km~s$^{-1}$, respectively. Thus
there is some disagreement in the velocities of the two features, but it
is within the uncertainties, which are due to possible errors in the
wavelength calibration, the uncertain relative contributions of the
components of the doublet, and the noise in the spectrum.  The uncertainty
in the wavelength scale is determined largely by the UKIRT spectrum, and
conservatively is 10~km~s$^{-1}$. The uncertainty in the centroid of the
doublet depends on the width and signal-to-noise ratio of the feature and
on the relative contributions of the two lines (separated by
36~km~s$^{-1}$, which is roughly 1:1 in Galactic dark clouds \citep{mcc99}
and toward the Galactic center \citep{oka05}. The uncertainty due to
feature's profile and signal-to-noise ratio is about 20 km~s$^{-1}$
whereas reasonable variation in the ratio of the doublet's components
translates into an uncertainty of about 5~km~s$^{-1}$ in the centroid.
Thus the overall uncertainty in the velocity centroid of the doublet is
about 30~km~s$^{-1}$. For the marginally detected singlet the uncertainty
in the centroid due to the noise is at least the point spacing which is
46~km~s$^{-1}$ (this can be seen by comparing the centroids of the UKIRT
2000 and Subaru 2002 spectra), giving an overall uncertainty in the
velocity centroid of roughly 50 km~s$^{-1}$ for that line.

Although the moderate discrepancy in velocities suggests the possibility
that the weak absorption at 3.932~$\mu$m is a noise fluctuation and that
the much stronger 3.882~$\mu$m absorption feature has an identification
other than \hhhp, we believe that this is unlikely. First, we have found
no other viable candidate for the 3.882~$\mu$m absorption other than
\hhhp. Second, in the numerous Galactic sources in which this doublet has
been detected there is no evidence of contamination of the doublet by
other lines. Third, we expect to detect \hhhp\ at roughly this strength
toward IRAS~08572+3915 based on the heavy obscuration of the nuclear
source and the evidence for a high column density of diffuse interstellar
gas and by analogy to the Galactic center. Finally, the radial velocity of
the 3.88~$\mu$m absorption, if due to \hhhp, is the same (to well within
the uncertainties) as that of the peak absorption by cold CO discussed in
the next subsection.

The equivalent widths of the two features are given in Table~2. Their
uncertainties are based on average point-to-point fluctuations in the
spectrum near the lines. The 3.88~$\mu$m doublet has more than three times
the equivalent width of the singlet. Using the standard equations relating
column density to equivalent width \citep{geb96}, we derive column
densities of 1.8~$\pm$~0.6~$\times$10$^{15}$~cm$^{-2}$ in the (1,1) para
level and 2.6$\pm$~0.6~$\times$10$^{15}$~cm$^{-2}$ in the (1,0) ortho
level. The most likely values yield a formal excitation temperature of
100~K, but this result is highly uncertain mainly due to the large
uncertainty in the equivalent width of the singlet. At densities typical
of molecular clouds the total \hhhp\ column density is the sum of the
above values, 4.4~$\times$10$^{15}$~cm$^{-2}$. If the temperature is
significantly higher than 100~K and the clouds are diffuse, a few higher
levels can be significantly populated and the total column density of
\hhhp\ could be higher, as shown by \citet{oka04}. In the Galactic Center,
where much of the \hhhp\ is in clouds at temperatures of $\sim$250~K, the
total column density of 4.3~$\times$10$^{15}$~cm$^{-2}$ is one-fourth
greater than the sum of the column densities in these lowest ortho and
para levels \citep{oka05}.

\subsection{CO}

Figure~3 shows the 4.90-5.05~$\mu$m spectrum of IRAS~08572+3915~NW
observed at a resolving power of 7500.  The spectrum is noisy and several
intervals within it are unrecoverable due to strong telluric absorption
lines. However, it is clear that the spectrum contains strong and broad
absorption lines of the fundamental band of CO, stretching across the
entire observed interval. Indeed the lines are so broad that in some
portions of the spectrum it is unclear if continuum gaps exist between
them.  The spacing of lines of the 1-0 band of CO is typically
550~km~s$^{-1}$. The apparent elevation of the continuum near 4.92~$\mu$m
might be due to H~I Pa~$\beta$ which is redshifted to 4.925~$\mu$m,
although the line appears to be considerably wider than that of shorter
wavelength infrared H~I recombination lines observed by \citet{gol95} and
\citet{vei99}.

The centroids of the CO lines in IRAS~08572+3915~NW are considerably
blueshifted from the central wavelengths determined from the systemic
redshift of the galaxy. Moreover, although the signal-to-noise ratio of
the spectrum is low, careful examination of Fig.~3 suggests that the lines
from the lower rotational levels are broader than those from high J. To
better test whether a separate velocity component is present in the low J
lines, we have coadded the spectra of the R(1), R(2), and P(1) lines as
well as those of the P(6), P(8), and P(11) lines, lines least affected by
telluric absorption. These average low J and high J spectra are compared
in Fig.~4. The low J line profiles have two strong absorption maxima, at
approximately -50~$\pm$25 and -150~$\pm$25~km~s$^{-1}$ relative to the
systemic velocity, whereas only a single strong absorption maximum, at
-160~$\pm$25 km~s$^{-1}$, is present for the high J profiles. Thus,
roughly speaking there is a warm blueshifted CO absorption component
extending roughly from 0 to -350 km~s$^{-1}$ and a cold component centered
near 0~km~s$^{-1}$, at approximately the same radial velocity as the
\hhhp\ lines. A warm redshifted component, that appears to be present in
the mean high J profile in Fig.~4, has recently been confirmed by
Shirahata et al. (2006). In the UKIRT spectrum it is centered at
+100~$\pm$30~km~s$^{-1}$ and is considerably weaker and narrower than the
other two components.

There is little or no evidence for CO absorption or emission from
vibrationally excited states. The v=2-1 R(6) transition (4.939~$\mu$m),
which occurs at the CO v=1-0 band center corresponds to a marginal
depression in the continuum between the strong CO 1-0 lines. However, the
continuum levels between the 1-0 lines at 5.02-5.06~$\mu$m, where the
v=2-1 line wavelengths lie midway between the 1-0 lines, are comparable to
the levels between 1-0 lines at 4.96-4.99~$\mu$m, where the 2-1 line
wavelengths coincide with the 1-0 lines.

The characteristic CO temperatures and column densities are difficult to
determine accurately from the narrow spectral range observed. Strong
blueshifted absorption lines are present at all observed J levels (up to
12), implying that the warm blueshifted CO component is close to or in
LTE, and that either the blueshifted portions of the medium J lines are
optically thick or the blueshifted gas has a range of temperatures. It
appears that the line from the highest J level observed, the P(12) line,
is somewhat weaker than the lines from lower levels. This and the
estimated strength of the R(0) and P(1) blue components allow us to
crudely and tentatively estimate the mean temperature of the warm
blueshifted component to be 200~(+100, -50)~K. Insufficient information is
available to estimate the temperature of the weak redshifted component,
but it may be warmer than the blueshifted component.

Assuming an isothermal optically thin slab in this temperature range, the
column density of the warm blueshifted component is
2~$\times$~10$^{18}$~cm$^{-2}$, with an uncertainty of a factor of two.
However, it is amost certainly an oversimplification to regard the warm
absorbing gas as isothermal and optically thin at all velocities. As
discussed in section 3.2, the profile is probably made up of spatially
separated velocity components. If each of these is absorbing along a
different line of sight to a different location on the continuum source,
the actual line optical depths are considerably greater than in the above
estimate. A column density an order of magnitude higher than the above
value would not be surprising.

This argument may also apply to the column density of the cold low
velocity absorption component.  Based on its strength and small number of
rotational levels that are populated, the column density of this component
is probably several times less than that of the warm blueshifted
component. The column density of the warm redshifted component is even
less. Assuming [CO]/[H$_2$]~=~1.5~$\times$~10$^{-4}$ \citep{lee96} the
lower limit on the hydrogen column density associated with the observed CO
(assuming the lines are optically thin) is
$N$(H$_2$)~$\approx$~1.5~$\times$~10$^{22}$~cm$^{-2}$. By comparison, from
the peak optical depth of 4.2 in the 10~$\mu$m silicate absorption feature
\citep{spo06a} and using $A_{\rm V}$/$\tau$$_{\rm sil}$~=~17.5
\citep{roc84, rie85} and $N$$_{\rm
H}$~=~1.9~$\times$10$^{21}$~cm$^{-2}$~mag$^{-1}$, we obtain
$N$(H$_2$)~=~7~$\times$~10$^{22}$~cm$^{-2}$, which is also a lower limit
because of the possibility of radiative transfer effects and/or foreground
silicate emission.

\section{Discussion}

\subsection{Low velocity \hhhp\ and CO}

The velocity of the \hhhp\ is close to the systemic velocity of
IRAS~08572+3915~NW, as indicated in Fig.~2. Although the absorption lines
of \hhhp\ appear fairly simple in shape, the low signal-to-noise ratio and
low resolution could be masking many details. Figure~2 includes the
spectrum toward the Galactic center source GCS3-2 recently observed at
6~km~s$^{-1}$ resolution by \citet{oka05}, but shown here with the same
binning as the IRAS~08572+3915~NW spectra. \citet{oka05} found that the
spectrally resolved profile of GCS3-2 has a full width at zero intensity
(FWZI) of 150~km~s$^{-1}$ and at least six discrete velocity components,
from both diffuse and dark clouds, some within a few tens of parsecs of
the center and some in distant spiral arms. The \hhhp\ absorption line
profiles towards IRAS~08572+3915~NW could be similarly complex.

The presence of a strong 3.4~$\mu$m interstellar feature suggests that a
considerable fraction of the interstellar molecular gas along the line of
sight toward the nuclear continuum source of IRAS~08572+3915~NW is in
diffuse clouds, as is the case towards the Galactic center \citep{whi97}.
If so and assuming that the total column density of \hhhp\ is not greatly
different from the sum of the column densities in the lowest ortho and
para levels, 4.4~$\times$~10$^{15}$~cm$^{2}$, we can estimate the column
length of absorbing \hhhp\ from its steady state density in diffuse
clouds. The value of $n$(\hhhp) depends on the assumed cosmic ray
ionization rate, $\zeta$. Assuming the previously canonical value of
$\sim$1~$\times$10$^{-7}$~cm$^{-3}$ for $n$(\hhhp) in diffuse clouds
\citep{mcc98, geb99} obtained if $\zeta$~=~3~$\times$10$^{-17}$~s$^{-1}$,
the derived length of the column containing \hhhp\ in IRAS~08572+3915~NW
is $\sim$~10~kpc, which clearly is unrealistically long. A similar
unreasonably high result is obtained for the Galactic center
\citep{geb99}. Using the 40 times higher cosmic ray ionization rate found
in one galactic diffuse cloud, but suspected of existing in many or all
such clouds \citep{mcc03} reduces the pathlength to a still high, but more
reasonable 300~pc. In the Galactic center \citet{oka05} have found that
the ionization rate is even several times higher than this value. If
conditions in IRAS~08572+3915~NW are similar, the pathlength then would be
reduced to less than 100~pc. If instead, all of the \hhhp\ were found in
dense clouds the pathlength though them would be $\sim$10~pc, but it then
would be difficult to explain the absence of \hhhp\ in the diffuse clouds
that produce the strong 3.4~$\mu$m feature. Moreover, the extinction
through the dense cloud material to the continuum source would be
unrealistically high, $\sim$1000~mag, based on the rough Galactic ratio
$N$(\hhhp) to $A_V$ of 3.6~$\times$10$^{12}$~cm$^{-2}$~mag$^{-1}$ as
reported by \citet{oka05}. Thus we conclude that the bulk of the observed
\hhhp\ is located in diffuse molecular gas.

The absorption lines of \hhhp\ and the cold CO have similar velocity
ranges. However, as in the case of the Galactic center \citep{oka05}, they
probably do not arise in all of the the same clouds. The CO abundance in
diffuse clouds is only one percent of the carbon abundance, whereas in
dense clouds almost all carbon is in CO. Thus the absorption by cold CO
must occur primarily in dense clouds, whereas, as argued above, the lines
of \hhhp\ are probably formed mainly in diffuse clouds. The present CO
data do not allow much to be inferred about the dense clouds, although
they do suggest from the width of the cold component that, as in the case
of the Galactic center, there are several of them along the line of sight.

\subsection{High velocity CO}

The discovery of high velocity CO features is the most striking result of
this paper.  Clearly there is no physical relation whatsoever between the
dominant blueshifted CO absorption and the \hhhp. This is because (1)
highly blueshifted velocities are not seen in the \hhhp\ lines, despite
the large CO column density in that component, and (2) the column
density of blueshifted \hhhp\ must be very low, because the length of the
column of blueshifted CO is very short, as discussed below.

The existence in the high velocity gas of strong CO lines from levels as
high as J=12 implies that the CO rotational population is maintained in
LTE to at least that level. Using our crude determination of 200~K for the
kinetic temperature and equating the Einstein A coefficient for the
J=12-11 transition (equation 1 in \citet{tho73}) to the collisional
excitation rate assuming a cross section for collisional excitation by
H$_{2}$ of 1~$\times$~10$^{-15}$~cm$^{-2}$ \citep[e.g.,][]{mck82}, we
estimate that the gas density $n$(\hh) must be at least
3~$\times$~10$^{6}$~cm$^{-3}$. For [CO]/[\hh] = 1.5~$\times$10$^{-4}$
\citep{lee96}, the lower limit to the observed warm CO column density of
2~$\times$~10$^{18}$~cm$^{-2}$ implies that the overall column length
of warm CO is less than than 0.001~pc, or less than 0.01~pc for an
order of magnitude higher CO column density.  The \hhhp\ column density
expected in such short column lengths would be undetectable.

Confinement of the observed wide range of blueshifted velocities to a
single clump of gas that is this thin and is detached from the continuum
source is inconsistent with the observations. Collisions between portions
of the gas in the clump at relative velocities of more than ten kilometers
per second would heat the gas significantly above the observed
temperatures, and collisions at more than several tens of kilometers per
second would dissociate even the CO. Yet there is no evidence in our
spectra for the CO being vibrationally excited. Thus if the absorbing gas
is detached it must be composed of numerous well separated thin sheets of
dense gas each of which contains a narrow range of velocities.

It is simpler to account for the wide range of blueshifted velocities if
the line-forming CO is located on the outer surface of an expanding,
more-or-less spherical, optically thick continuum source. In such a
geometry a range of absorption velocities would be present, from near zero
near the edges of the source to most highly blueshifted at the center.
This simple model implies an expansion velocity of $\sim$350~km~s$^{-1}$
and a column much higher than the upper in section 2.2, as discussed
earlier.

This model cannot account for the weak redshifted CO absorption component
seen at +100~km~s$^{-1}$, however.  Judging from the relative strengths of
the low and high J absorptions at this velocity (see Fig.~4), the
redshifted CO appears is warmer than the blueshifted CO, and therefore
probably is located interior to it. Thus the actual geometry of the
continuum source and the absorbing gas may be quite complex.

\subsection{Star Formation or AGN}

It has been argued by \citet{eva02} that star formation is rampant in the
central regions of IRAS~08572+3915~NW, and thus we consider if the
observed high velocities of the CO lines could be the result of such
activity. The mid-infrared continuum of IRAS~08572+3915~NW arises in a
region less than 250~pc in diameter \citep{soi02} and thus the putative
young stars would be confined to that region, which is roughly the size of
the CO millimeter line emission \citep{eva02}. In star-forming regions
winds from young stellar objects (YSOs) shock-heat and sweep up cloud
material. \citet{vei99} have detected emission lines in IRAS~08572+3915~NW
from the first excited vibrational state of \hh, which could be emitted by
shock-heated gas as a result of star formation. The observed absorption by
high velocity CO could arise in numerous dense and thin shells of swept-up
post-shock gas in a large number of discrete star-forming clouds seen
against the continua from the a widely distributed set of YSOs. A large
preponderance of blueshifted absorption over redshifted absorption would
be expected, as is observed.

Despite the above considerations this explanation of the observed
continuum and high velocity absorption lines as being the result of
myriads of individual star-forming events over a very extended region
seems contrived. Moreover, the mean velocity of the warm blueshifted CO is
considerably higher than would be expected based on observations of
outflows from YSOs in the Galaxy. A further difficulty with it is that,
even in the most active star-forming regions in the Galaxy, such as OMC-1,
the bulk of the molecular gas is at temperatures much less than 200~K.
Only a small portion of the gas, located just downstream from the shocks,
is so warm. However, in IRAS~08572+3915~NW the bulk of the gas producing
the absorption lines is at this warm temperature.

Thus we suspect that events associated with star formation are unlikely
to be responsible for the CO line profiles. Adaptive optics imaging on a
large telescope of the infrared K-band continuum might provide important
information on its spatial distribution and a way of discriminating
between domination of the energetics by an extended starburst or a more
compact source of radiation.

\citet{ima01} have claimed that a buried AGN must be the dominant
luminosity source of IRAS~08572+3915~NW (and some other ULIRGS).  In that
case one would expect the infrared continuum source to be compact. The
preponderance of blueshifted warm gas in front of the nucleus of
IRAS~08572+3915~NW naturally suggests ejection of material from the AGN.
The luminosity of the central source is 2~$\times$~10$^{12}$~L$_{\odot}$
\citep{dud97}; hence the $\sim$200~K CO must be $\sim$10~pc from the AGN.
This size and the observed expansion speed imply that the ejection began
$\sim$30,000 years ago. If the cloud forms a complete shell around the
AGN, the mass of the ejected material exceeds 2000~M$_{\odot}$, and the
kinetic energy exceeds 3~$\times$~10$^{50}$~ergs, roughly comparable to
the total energy liberated in a supernova.

In the unified model of AGN, the differences between Type 1 and Type 2
objects is explained by invoking a rotating toroidal cloud of dust and gas
that obscures gas in the broad line region from some viewing angles while
leaving it exposed from others. The presence of both blueshifted and
redshifted CO absorptions in IRAS~08572+3915~NW suggests the possibility
of rotation of the absorbing gas about the central luminosity source. As
in the case of expansion, the absorbing CO would need to be situated on
the outside surface of the dust in order for red and blueshifted
components to be seen in absorption against it. The major problem with
this model is the extreme weakness of the redshifted absorption relative
to the blueshifted absorption. It implies a highly asymmetric distribution
of rotating material, a highly asymmetric distribution of continuum
emission, or variation of the foreground extinction by several tens of
visual magnitudes across the source of the infrared continuum. The
possibility that the redshifted gas is at a higher temperature than the
blueshifted gas is a second potential difficulty.

Thus one cannot easily interpret our observations as arising in a
torus-like gaseous structure. We tentatively interpret them as probing a
transient event (occurring much more rapidly than the galaxy-galaxy
interaction) in the nucleus of IRAS~08572+3915~NW. Neither the
consequences of a starburst or "steady-state" phenomena related to an
obscured AGN (e.g., a stable torus) can provides an explanation for the CO
observations. Although the transient event appears to be mainly one of
ejection of material from the nucleus, the simultaneous observation of a
small amount of material approaching the nucleus suggests both a complex
geometry and complex gas motions.

\section{Conclusion}

The broad absorption lines of CO toward IRAS~08572+3915~NW consitute the
most remarkable finding of this paper. The highly velocity-shifted and
warm CO implies a transient event in the nucleus, more likely to be
associated with AGN activity than with massive star formation. No
observations of IRAS~08572+3915~NW directly related to this phenomenon
appear in previously published papers.  Warm CO is not unique to this
ULIRG, however. \citet{spo04} have found broad absorption due to the
fundamental band of CO in IRAS~F10183-7111 and have recently reported
several similar detections in other deeply obscured galactic nuclei
\citep{spo06b}.  All of their observations were made at much lower
spectral resolution, and hence do not resolve individual lines and
determine velocities. Thus it is unclear if the apparently energetic
events taking place in IRAS~08572+3915~NW are also occurring in the other
galaxies.  In some cases, e.g. IRAS~F10183-7111, the CO that Spoon et al.
have found is considerably warmer and has a larger column density than
IRAS~08572+3915~NW, but their general conclusions about the shortness of
the absorbing column and proximity to the central luminosity source are
similar to ours.

The absorption lines of \hhhp\ toward IRAS~08572+3915 probably arise
largely in diffuse gas, as do those seen toward the Galactic center.  In
combination with the low velocity low J lines of CO, the \hhhp\ could in
principle provide much more detailed information on the nature of the
interstellar medium of IRAS~08572+3915~NW. However, considerably higher
resolution and higher sensitivity measurements than those reported here
are required. Sensitive measurements of the R(2,2)$^{l}$ and metastable
R(3,3)$^{l}$ line could much more tightly constrain the density and
temperature of the \hhhp-containing clouds, as they have done for the
\hhhp\ seen along the line of sight to the Galactic center \citep{oka05}.

The detection of extragalactic \hhhp\ and the high resolution infrared
spectra of extragalactic CO reported here and by \citet{spo03}, which have
resolved the fundamental band into individual lines, are the first
measurements of their types. In the future, using ground-based 8--10~m
class and larger telescopes along with the James Webb Space Telescope, one
can anticipate that infrared spectroscopy of interstellar CO and \hhhp, in
combination with measurements of other molecules and dust, will be a
standard technique for probing the diffuse and dense clouds and gauging
the nuclear activity of many distant galaxies.

\begin{acknowledgements}

We thank the staffs of UKIRT and Subaru for their support.  We are
grateful to F. Lahuis, M. Shirahata, H. Spoon, and the referee for a
number of helpful comments.  T. R. G.'s research is supported by the
Gemini Observatory, which is operated by the Association of Universities
for Research in Astronomy, Inc., on behalf of the international Gemini
partnership of Argentina, Australia, Brazil, Canada, Chile, the United
Kingdom and the United States of America. M. G. is supported by a
Fellowship for Research Abroad from the Japan Society for the Promotion of
Science. T. O. is supported by NSF grant PHY 03-54200.

\clearpage

\end{acknowledgements}

\clearpage

\begin{figure} 
\epsscale{1} 
\plotone{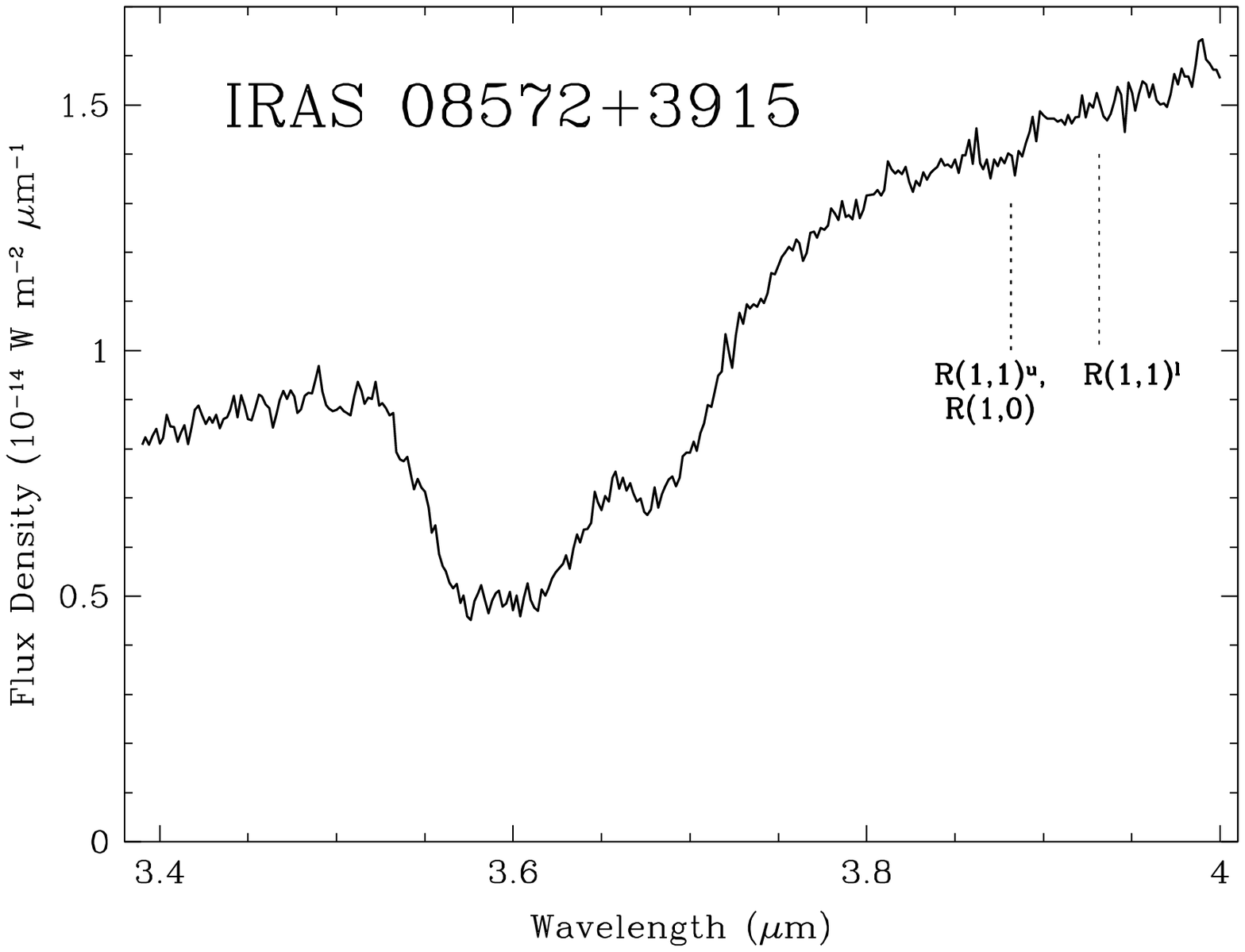} 
\caption{The 3.4-4.0~$\mu$m
spectrum of IRAS~08572+3915~NW at R$\sim$1500 showing the redshifted
3.4~$\mu$m interstellar hydrocarbon band and the locations of the lines of
\hhhp\ detected at higher resolution. The noise can be estimated by the
fluctuations in $\sim$0.05~$\mu$m wide intervals.}
\label{fig1}
\end{figure}

\clearpage

\begin{figure} \epsscale{1} \plotone{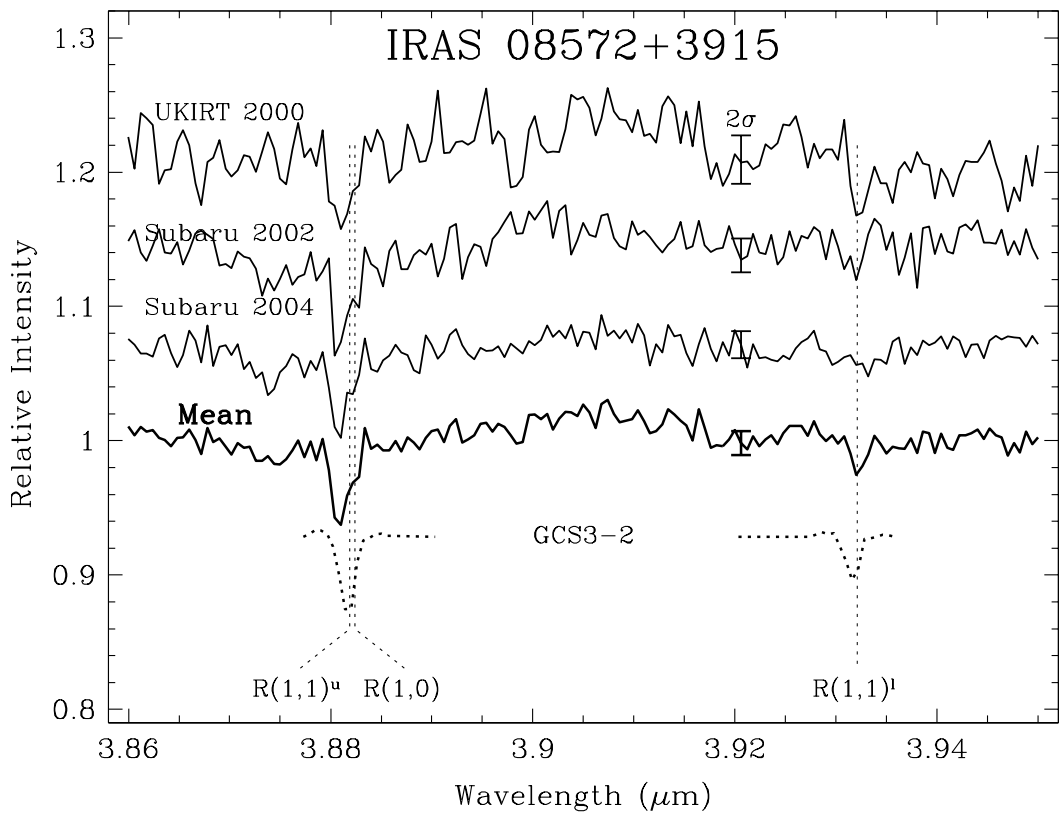} \caption{Individual and
mean spectra from UKIRT and Subaru of IRAS~08572+3915~NW near the
locations of three lines of \hhhp. The high resolution spectrum of the
R(1,1)$^{l}$ line in the Galactic center source GCS3-2 \citep{oka05},
binned to the same point-to-point spacing as the spectra of
IRAS~08572+3915~NW, is shown at lower right and also was used to make a
model spectrum of the 3.88~$\mu$m doublet at lower left.}
\label{fig2}
\end{figure}

\clearpage

\begin{figure} \epsscale{1} 
\plotone{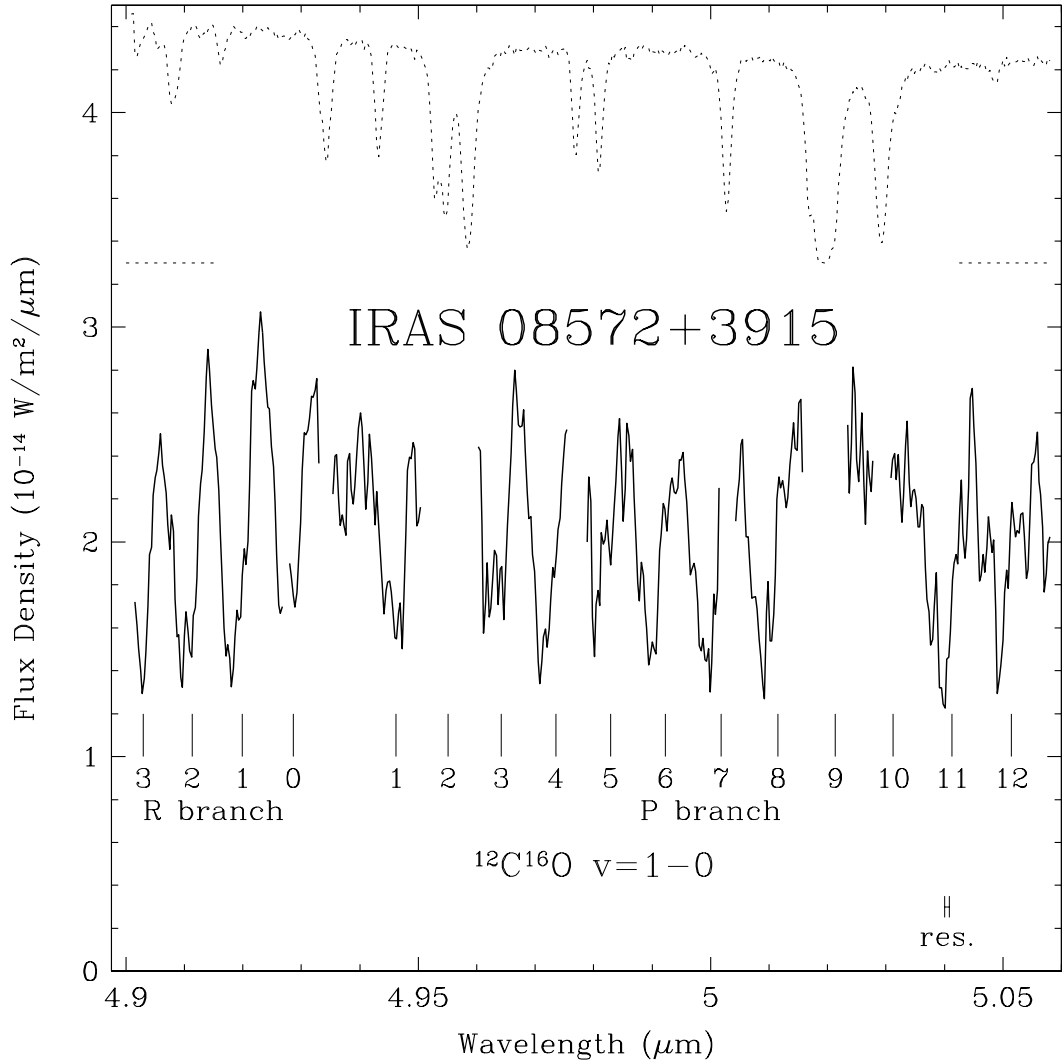} 
\caption{Spectrum of
IRAS~08572+3915~NW at R=7500 covering a portion of the fundamental band of
CO. The atmospheric transmission in shown as a dashed line. CO v=1-0 band
line positions are indicated at the bottom, for a redshift of 0.05821.}
\label{fig3}
\end{figure}

\clearpage

\begin{figure}
\epsscale{1}
\plotone{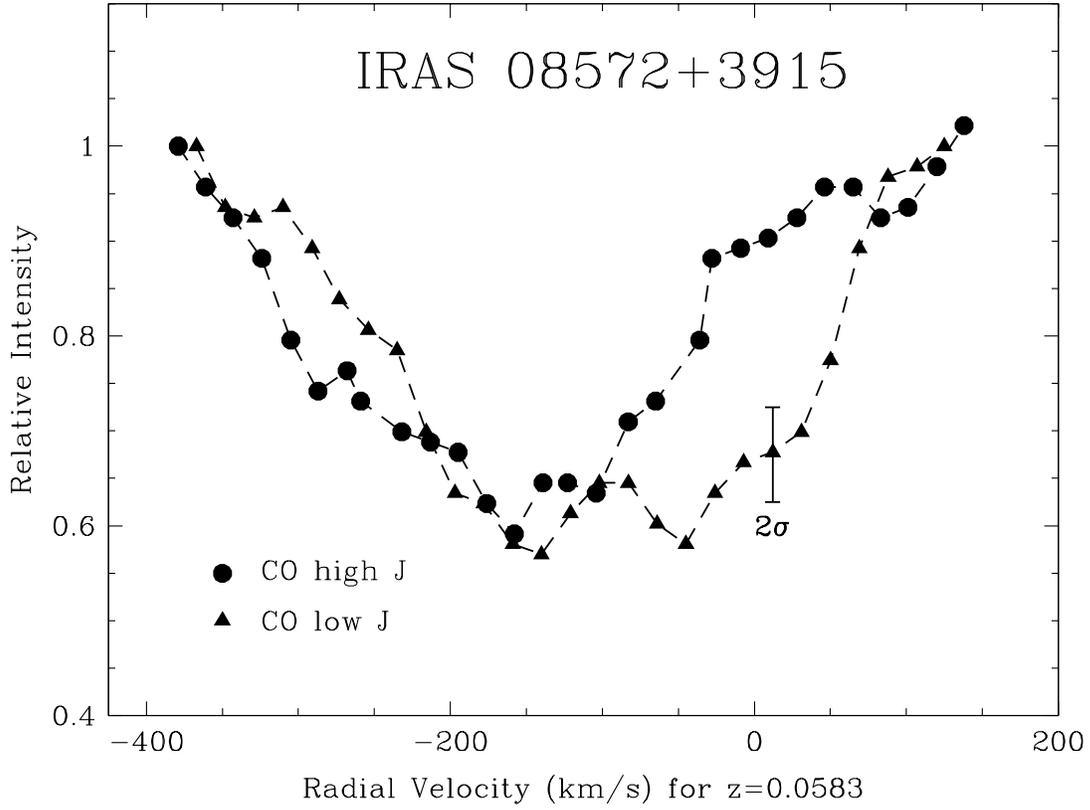}
\caption{Velocity profiles of CO lines from low and high J levels. Circles
are the mean of the 1-0 R(1), R(2), and P(1) lines; triangles indicate
the mean of the P(6), P(8) and P(11) lines. A representative error bar
($\pm$1$\sigma$ is shown.}
\label{fig4}
\end{figure}

\clearpage

\begin{deluxetable}{ccccccc}
\small 
\tablecaption{Observing Log}
\tablehead{
\colhead{UT Date} &
\colhead{Telescope} &
\colhead{Wavelength} &  
\colhead{R} &
\colhead{Exposure} &
\colhead{Weather} &
\colhead{Calib. star} \\
 & & \colhead{($\mu$m)} & &\colhead{(min)} & & }
\startdata
19981227 & UKIRT  & 3.39-4.00   &  1500 &   78 & clear  & HR 3690 (A3V)\\
20001208 & UKIRT  & 3.82-3.99   &  6000 &  130 & clear  & HR 2818 (A1V)\\
20010109 & UKIRT  & 4.89-5.06   &  7500 &   90 & clear  & HR 3579 (F5V)\\
20020224 & Subaru & 3.85-3.96   & 10000 &   67 & clear  & HD 90470
(A2V)\tablenotemark{a}\\
20040209 & Subaru & 3.85-3.96   &  5000 &   80  & clouds & HR 2891 (A1V)\\
20040402 & Subaru & 3.85-3.96   & 10000 &   92 & clear  & HR 2891 (A1V)\\
\enddata
\tablenotetext{a}{Calibration star observed on different date due to rapid
change in weather}
\end{deluxetable}

\clearpage

\begin{deluxetable}{ccc}
\tablecaption{\hhhp\ Line Equivalent Widths}
\tablehead{
\colhead{Feature} &
\colhead{Wavelength ($\mu$m)} &
\colhead{W$_{\lambda}$} 
}
\startdata
R(1,1)$^{u}$ + R(1,0) & 3.8813 & 1.44$\pm$0.14~$\times$~10$^{-4}$\\
R(1,1)$^{l}$          & 3.9322 & 0.39$\pm$0.12~$\times$~10$^{-4}$\\
\enddata
\end{deluxetable}

\end{document}